\documentclass[9pt, sigconf]{acmart}

\usepackage{color}
\usepackage{graphicx}
\usepackage{amsmath}
\usepackage{graphicx}
\usepackage{soul}

\usepackage{subcaption}
\usepackage{caption}
\usepackage[utf8]{inputenc}

\usepackage{tikz}

\makeatletter
\def\hlinewd#1{%
\noalign{\ifnum0=`}\fi\hrule \@height #1 %
\futurelet\reserved@a\@xhline}
\makeatother
\usepackage{booktabs}
\usepackage{multirow}

\usepackage[binary-units=true]{siunitx}

\usepackage[algoruled,noline,longend,linesnumbered]{algorithm2e}

\usepackage{pgfplots}
\usepackage{pgfplotstable}
\usepackage{siunitx}
\usepackage{algorithm2e,amsmath}

\SetKwInOut{Input}{Input}
\SetKwInOut{Output}{Output\,}
\SetKwInOut{Data}{Data}
\SetKwProg{Tree}{Tree}{}{EndTree}

\makeatletter

\usepackage{amsthm}

\usepackage[small,compact]{titlesec} 
\usepackage[compact]{titlesec}

\setcopyright{rightsretained}
\renewcommand\footnotetextcopyrightpermission[1]{} 

\acmISBN{978-1-4503-XXXX-X/18/06}

\begin{document}

\graphicspath{{Fig/}}
\def\figname{Figure}
\def\algname{Algorithm}

\begin{abstract}
Deep neural networks (DNNs) have made breakthroughs in various fields including image recognition and language processing. DNNs execute hundreds of millions of multiply-and-accumulate (MAC) operations. To efficiently accelerate such computations, analog in-memory-computing platforms have emerged leveraging emerging devices such as resistive RAM (RRAM). However, such accelerators face the hurdle of being required to have sufficient on-chip crossbars to hold all the weights of a DNN. Otherwise, RRAM cells in the crossbars need to be reprogramed to process further layers, which causes huge time/energy overhead due to the extremely slow writing and verification of the RRAM cells. As a result, it is still not possible to deploy such accelerators to process large-scale DNNs in industry. To address this problem, we propose the BasisN framework to accelerate DNNs on any number of available crossbars without reprogramming. BasisN introduces a novel representation of the kernels in DNN layers as combinations of global basis vectors shared between all layers with quantized coefficients. These basis vectors are written to crossbars only once and used for the computations of all layers with marginal hardware modification. BasisN also provides a novel training approach to enhance computation parallelization with the global basis vectors and optimize the coefficients to construct the kernels. Experimental results demonstrate that cycles per inference and energy-delay product were reduced to below 1\% compared with applying reprogramming on crossbars in processing large-scale DNNs such as DenseNet and ResNet on ImageNet and CIFAR100 datasets, while the training and hardware costs are negligible.

\end{abstract}

\newcommand{\papertitle}{BasisN: Reprogramming-Free RRAM-Based In-Memory-Computing by Basis Combination for Deep Neural Networks}

\title{\papertitle}
%


\title{\papertitle}
\author{ Amro Eldebiky$^1$, Grace Li Zhang$^2$, Xunzhao Yin$^3$, Cheng Zhuo$^3$, Ing-Chao Lin$^4$, Ulf Schlichtmann$^1$, Bing Li$^5$\\
$^1$Technical University of Munich, $^2$Technical University of Darmstadt, $^3$Zhejiang University, $^4$National Cheng Kung University, $^5$University of Siegen\\
Email: \{amro.eldebiky, ulf.schlichtmann\}@tum.de, grace.zhang@tu-darmstadt.de, iclin@mail.ncku.edu.tw, bing.li@uni-siegen.de\\
}


\maketitle
\pagestyle{plain}



\section{Introduction} \label{sec:intro}
Deep neural networks (DNNs) have been successfully utilized across various domains, such as image recognition \cite{krizhevsky2012imagenet} and language processing \cite{chiu2018state}. The effectiveness of DNNs in achieving high accuracy is attributed to the extensive use of multiple layers \cite{deeplearning}, resulting in a substantial number of weights and multiply-and-accumulate (MAC) operations within DNNs. 
To accelerate DNNs, analog in-memory-computing (IMC) platforms leveraging emerging technologies such as resistive RAM (RRAM) \cite{chi2016prime,wan2022compute,8714954,9116244,9073995}, optical components \cite{9256423,10323877} and Ferroelectric FET (FeFET) \cite{9474234} have been introduced. Among them, RRAM-based accelerators demonstrate promising energy efficiency.

RRAM-based IMC accelerators, so far, follow a weight-stationary approach to execute DNNs.
In this approach, RRAM cells need to be programmed to target conductances to represent weights of a DNN. 
In this way, RRAM cells store the weights of a DNN. 
The multiplication operations in the DNN can then be executed by applying voltages on such cells. The resulting currents are accumulated to realize addition operations. Accordingly, RRAM-based IMC platforms implement MAC operations based on Ohm’s law and Kirchhoff’s current law, so that their computational and energy efficiency is very high.



RRAM-based IMC platforms, however, suffer from critical issues which hinder their practical application in executing DNNs. 
One of the issues is the time-consuming programming process of RRAM cells to perform inference of DNNs. 
For example, the number of cycles needed by the novel RRAM programming approaches to reprogram a $128 \times 128$ RRAM crossbar is $10^4\sim 10^5$ cycles \cite{chen2023novel,merced2016repeatable}. 
Another issue is the limited crossbar size and the limited number of crossbars available on-chip to store all the weights of a DNN, so that reprogramming the RRAM cells is required to reuse the crossbars. For example, \cite{wan2022compute} manufactured an RRAM-based IMC chip with 48 crossbars of dimension $256 \times 256$, which is not sufficient to execute DNNs such as DenseNet, ResNet and large language models (LLMs). To execute such DNNs, the computations have to be halted to wait for the slow reprogramming process.

Previous work tried to tackle such problem by two different approaches. The first approach is trying to reduce the programming/reprogramming time. 
 For example, \cite{merced2016repeatable} introduces a method to program the RRAM crossbars elements in a row-wise approach rather than element by element. \cite{chen2023novel} further proposes a block-based reprogramming method with a multi-row programming
algorithm.  
The second approach is trying to compress the DNNs in a way that matches the size and the number of the available RRAM crossbars. \cite{wang2023epim} represents neural network operations with reduced-size parameters called epitomes to compress DNNs. \cite{chu2020pim} uses fine-grained pruning to compress weight matrices in DNNs that fit the crossbar
size to reduce the demand of crossbars.

However, such approaches marginally address the problem but do not eliminate the necessity of reprogramming in crossbars for large-scale DNNs. The reprogramming time in \cite{merced2016repeatable, chen2023novel} still causes significant slowdown of the inference process when reprogramming is needed in an RRAM-based IMC accelerator for large-scale DNNs.  
Besides, the existing compression ratios in \cite{wang2023epim, chu2020pim} are not sufficient to compress backbone DNNs such as ResNet and DenseNet to be deployed on the available RRAM-based IMC accelerators without the need of reprogramming.

Different from the approaches above, in this paper, we introduce BasisN, a method to avoid the requirement of reprogramming RRAM crossbars for large DNNs by representing the layers' kernels of DNNs as combinations of a basis system. The key contributions of this work are as follows:
\begin{itemize}
        \item BasisN suggests a novel kernel representation in RRAM-based IMC accelerators. The kernels of all the layers of a DNN are represented as combinations of a set of global basis vectors which are written to RRAM crossbars only once.
       
       \item The BasisN training framework trains DNNs such that all weight matrices are combination of basis vectors that have been initially
    written in crossbars, while 
      minimal bitwidth is required for the coefficients combining the basis vectors. 
       

    \item The introduced new technique fits large DNNs on any number of available crossbars without reprogramming the crossbars while requiring much fewer computational cycles for inference and a minimal hardware overhead for the newly introduced representation.
    
    \item Experimental results demonstrate that the number of cycles per inference and energy-delay product can be reduced to less than 1\% compared with the state-of-the-art approaches with reprogramming \cite{merced2016repeatable, chen2023novel}, while  no degradation of the inference accuracy and only negligible hardware cost are incurred.
    
\end{itemize}

The rest of this paper is organized as follows. In
Section~\ref{sec:preliminaries}, the background and the motivation of
this work are explained. In Section~\ref{sec:methods}, we introduce the new BasisN
concept including a training framework and the corresponding hardware architecture.  Experimental results are reported in
Section~\ref{sec:results} and
conclusions are drawn in Section~\ref{sec:conclusion}.

\section{Motivation}
\begin{figure}
    \centering
    \includegraphics[width=0.97\linewidth]{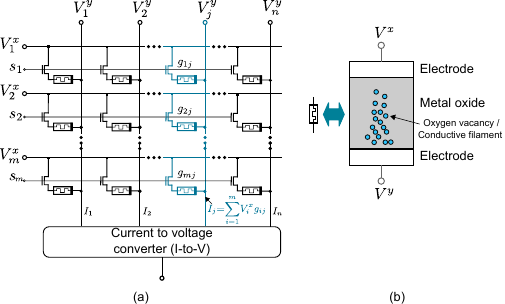}
    \caption{(a) The structure of an RRAM crossbar. (b) The structure of an RRAM cell.}
    \label{fig:crossbar}
\end{figure}

\label{sec:preliminaries} 
RRAM-based IMC platforms take advantage of analog computing to enhance the 
computational and energy efficiency in executing DNNs. 
Figure~\ref{fig:crossbar} depicts an RRAM-based crossbar structure, where RRAM cells are positioned at the intersections of wordlines and bitlines. Transistors are employed to activate RRAM cells. In order to execute multiply-accumulate (MAC) operations, RRAM cells are initially programmed into target  conductance values to represent weights of a DNN. Subsequently, voltages are applied to the horizontal wordlines while the vertical bitlines are connected to ground. This process leads to a current flowing within each RRAM cell, resulting in the multiplication of the cell's conductance value and the applied voltage. The accumulated currents at the bottom of each column represent the addition results.

Due to the high computational and energy efficiency, RRAM crossbars have been deployed to be a general-purpose hardware accelerator to execute the inference of various DNNs. 
Under a given area constraint, however, the number of such crossbars and the crossbar size are limited. For example, \cite{wan2022compute} manufactured an RRAM IMC chip with 48 crossbars with a size of $256 \times 256$. 
Such crossbars can only store around $48\times 256 \times 256=3,145,728$ different conductances to represent weights, which is much smaller than the number of weights in DNNs such as DenseNet and ResNet. 
Accordingly, reprogramming is required to update the conductances of RRAM cells in the chip to fully execute all the MAC operations in such DNNs. 


The cost of programming/reprogramming RRAM cells accurately to target conductance values for computation is significantly high in terms of computational efficiency and energy. 
For example, the number of cycles needed by the novel RRAM programming approaches to reprogram a $128 \times 128$ RRAM crossbar is $10^4\sim 10^5$ cycles \cite{chen2023novel,merced2016repeatable}. Accordingly, this reprogramming process causes a significant slowdown of the whole chip. 

\begin{figure}
    \centering
    \includegraphics[width=\linewidth]{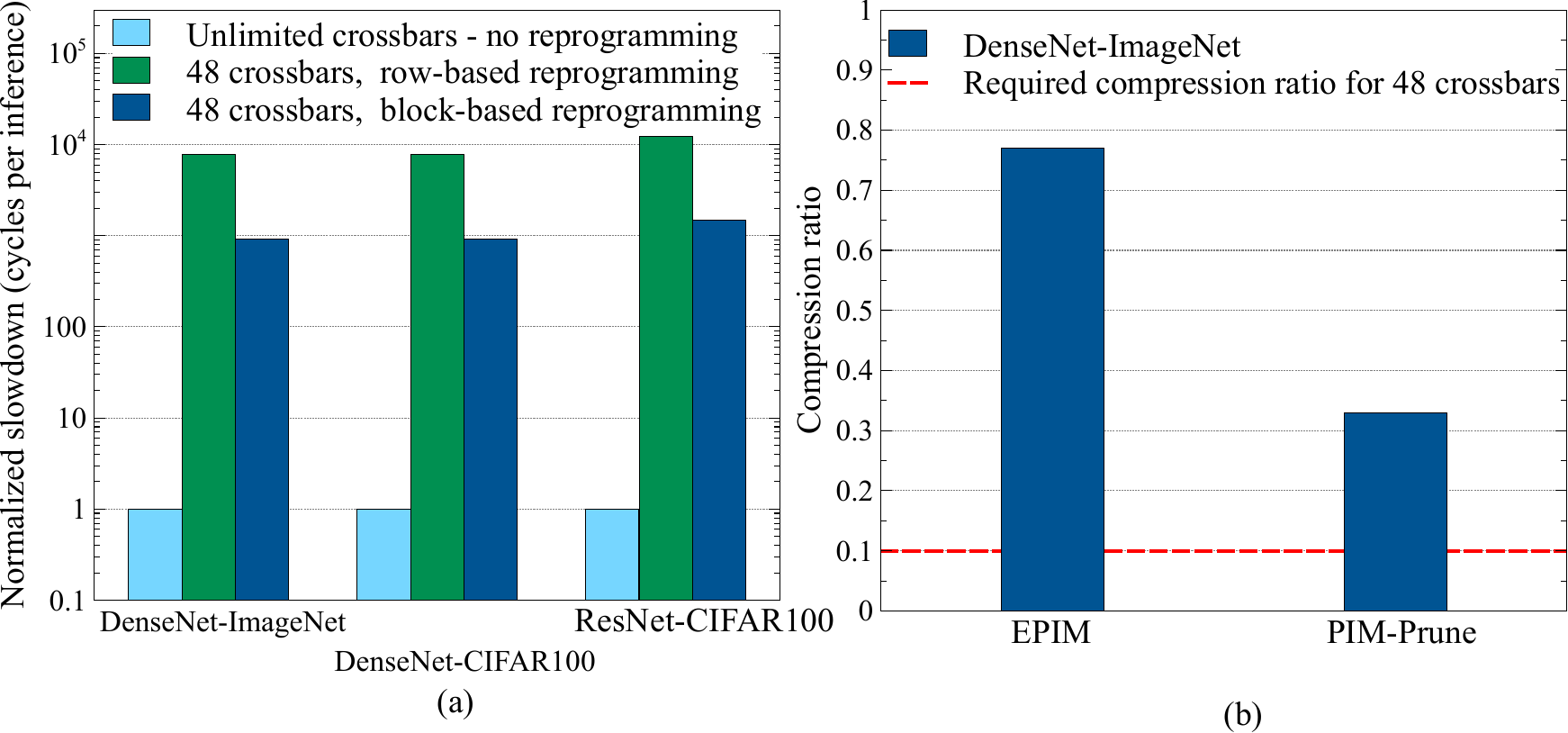}
    \caption{a) Performance slowdown due to reprogramming when several benchmarks are deployed on 48 RRAM crossbars of a size $256\times 256$, with row-based reprogramming \cite{merced2016repeatable} and block-based reprogramming \cite{chen2023novel}. b) The compression ratios achieved by EPIM \cite{wang2023epim} and PIM\_prune \cite{chu2020pim} for DenseNet-ImageNet benchmark and the required compression ratio to avoid reprogramming. }
    \label{fig:motivation2}
\end{figure}

To verify this performance slowdown, we evaluated the execution cycles per inference for three cases, e.g., no reprogramming with unlimited number of crossbars, row-based programming strategies \cite{merced2016repeatable} with 48 crossbars, and block-based programming strategies \cite{chen2023novel} with 48 crossbars. 
In the three cases, the crossbar size is $256 \times 256$. 
The results are shown in Figure~\ref{fig:motivation2}(a), where the x-axis represents the tested benchmarks, and the y-axis represents the ratio of inference cycles to the case with unlimited number of crossbars where no reprogramming is needed. According to this figure, a slowdown with factors $>>1000$ due to reprogramming is observed even with a fast block-based reprogramming approach. 



To address the challenge above, previous techniques also 
compress the DNNs in a way to 
reduce the number of required crossbars to avoid time-consuming reprogramming \cite{wang2023epim, chu2020pim}. However they cannot solve this problem completely.
To verify this, we evaluated the required compression ratio of DenseNet-ImageNet benchmark to fit the DNN on 48 crossbars with a size of $256\times 256$ without reprogramming. Figure~\ref{fig:motivation2} (b) shows the results. According to this figure, DenseNet-ImageNet would need a compression ratio to less than 0.1 to fit on the 48 crossbars. However, the compression approaches, EPIM in \cite{wang2023epim} and PIM\_Prune in \cite{chu2020pim} can not achieve such compression ratios. Accordingly, reprogramming of crossbars is inevitable in all previous approaches.

\section{The Proposed BasisN Framework}
\label{sec:methods}
The BasisN framework aims to eliminate the inability to deploy large DNNs on any RRAM IMC accelerator with any number of available crossbars without reprogramming. 
BasisN presents a new computing approach to have all the kernels in a DNN being trained as combinations of a basis system vectors of size N corresponding to the crossbar dimension with a limited number of allowed combinations coefficients. Accordingly, only the basis vectors need to be written in all the RRAM crossbars once. The computation of one kernel is then obtained as a combination of the individual multiplications between the inputs and the basis vectors written in the crossbar columns with the defined coefficients . Accordingly, any crossbar can be used for the computation of any kernel in any layer and the reprogramming of the crossbars is not needed at all.

In this section, the BasisN computation concept and steps are elaborated in detail showing the gains and the minimal hardware modifications needed to implement such approach. Besides, a novel training approach to determine the global basis vectors and the coefficients for the layers is presented. The BasisN training approach can deal with the two scenarios of either training a DNN from scratch or to start with a pretrained model to benefit from knowledge of large DNNs.

\subsection{BasisN kernel representation and hardware architecture}
\label{section:representation}

\begin{figure}[!t]
    \centering
    \includegraphics[scale=0.75]{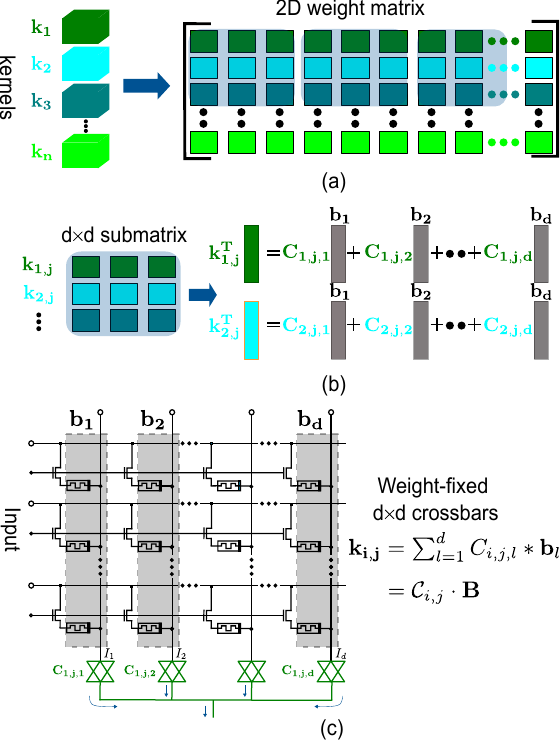}
    \caption{BasisN representation of the weights of a convolutional layer. a) The kernels of the layer, reshaping of the kernels as 2D weight matrix and partitioning into $d \times d$ submatrices fitting into the crossbars. b) The representation of a kernel partition as a combination of the basis vectors. c) The implementation of the BasisN representation on the crossbar hardware.}
    \label{fig:method}
\end{figure}

In weight-stationary approaches, the weights of any DNN layer are reshaped to a 2D-matrix. For example, Figure~\ref{fig:method}(a) shows a convolutional layer with weights of shape $(n,t,w,h)$. $n$ represents the number of kernels, $t,w,$ and $h$ represent the depth, the width, and the height of the kernel, respectively. Such weights are reshaped to a 2D matrix of shape $(n, t*w*h)$ in which each kernel is flattened and represented as a row. 
The 2D weight matrix is partitioned into submatrices matching the crossbar size $d\times d$ when $d<t*w*h$. The submatrices are then mapped to the crossbars. Each column in a crossbar represents one partition of a kernel/row in the 2D matrix. The partial results from the crossbars representing the MAC outputs of the submatrices are accumulated together to implement the complete MAC operation of a corresponding row.

Alternatively, BasisN proposes a novel kernel representation in crossbars to avoid reprogramming. A fundamental concept in linear algebra is the ability to represent any vector in a vector space as a linear combination of basis vectors spanning that space \cite{principles}. For a vector space of dimension m, denoted as $\mathbf{V^m}$, with a set of linearly independent basis vectors $\{\mathbf{v_1, v_2, ..., v_m}\}$, any vector $\mathbf{v}$ can be represented in terms of the basis as $\mathbf{v}=\sum^m_{i=1} c_i\mathbf{v_i}$, where $c_i \in \mathbb{R}$. BasisN exploits such concept to represent the kernels in DNNs. In this approach, a
set of basis vectors are pretrained and written into available crossbars, which is less than the number of needed crossbars to store all the weights of the DNN. Such basis vectors are used to reconstruct the kernels in the DNN with coefficients determined by the proposed method.
Only these coefficients need to be changed during computing, which are implemented by transmission gates at the bottom of the columns of the crossbars. 



A weight matrix of dimension $(n,t*w*h)$ is partitioned into submatrices matching the crossbar size $d\times d$. For example, the $d \times d$ submatrix in Figure~\ref{fig:method}(b) corresponds to the subkernels in the upper-left corners of the 2D weight matrix in Figure~\ref{fig:method}(a).
Each row in the submatrix represents a partition of a kernel $\mathbf{k_{i,j}}$, where $i$ is the kernel's index and $j$ is the partition's index.
In BasisN, such partition is represented as $\mathbf{k_{i,j}}=\sum^{d}_{l=1} C_{i,j,l}*\mathbf{b}_l $ where $\{\mathbf{b_1, b_2, ..., b_{d}}\}$ is the set of basis vectors hosted in crossbars and shared by all the subkernels of the DNN mapped onto the crossbars. $d$ is the crossbar dimension to partition the weight matrices representing the size of the vector space and basis system.
Written in a vector format, 
the subkernel $\mathbf{k_{i,j}}$ can be expressed as 
$\mathbf{k_{i,j}}=
[C_{i,j,1}\dots C_{i,j,d}]
\cdot[\mathbf{b_1, b_2, ..., b_{d}}]^T=\mathcal{C}_{i,j}\cdot \mathbf{B}$, where $\mathbf{B} $ is the matrix formed by the basis vectors. 
 The coefficients $C_{i,j,l}$ in $\mathcal{C}_{i,j}$ are limited to specific values or quantized to allow hardware-friendly time-multiplexed computation as shown in the next subsections.


To explain the proposed hardware architecture, we use the simplest case of having 1-bit control coefficients as shown in Figure~\ref{fig:method}(c).
In this case, kernel 
 $\mathbf{k_{i,j}}$ is implemented by the combination of the basis vectors 
 $\{\mathbf{b_1, b_2, ..., b_{d}}\}$ as 
 $\mathbf{k_{i,j}}=\sum^d_{l=1} C_{i,j,l}*\mathbf{b}_l$
 with $C_{i,j,l} \in \{0,1\}$.
%
At the bottom of each crossbar column, 
 a transmission gate (TG) is added and controlled by the binary control coefficients $C_{i,j,l}$ to implement the dot product of the input and one basis vector.  
 The outputs of the TGs are all connected together and the accumulated currents implement the controlled sum of the multiplications of the input with all basis vectors.
The TGs are controlled by single bits to select or deselect the corresponding column. 


\subsection{Multibit control coefficients over multicycles in BasisN}

In Figure~\ref{fig:method}, the simplest case of binary control coefficients is shown. To enhance the accuracy of representing the weight matrices with a limited number of crossbars, BasisN also allows coefficients $C_{i,j,l}$ to have multiple bits. The computations with these multiple bits are implemented using time multiplexing in BasisN. 
At each time step, the computation for one bit significance of the control coefficients is conducted similar to the single bit implementation explained above.
The partial results from the time steps are shifted according to their bit significance 
and accumulated in the output registers in the digital domain. 
For example, if the control coefficients are quantized to 4 bits, the computations are conducted over 4 time steps. During the first step, the lowest bits of all the coefficients are selected to control the TGs at the bottom of all the crossbars. In the following cycles, further bits in the coefficients are selected to control the TGs, and the results are shifted by 1, 2, and 3 bits, respectively, before they are accumulated to the first results, thus implementing the multiplication of the power of 2 corresponding to the bit locations in the coefficients.


In this multi-bit implementation, the more bits a coefficient has, the more accurate the basis vectors can be combined to implement the weight matrices. However, more cycles are needed to implement these bits, leading to a tradeoff between accuracy and performance.


\subsection{BasisN alternating training of basis and coefficients}
\label{sec_training}
In BasisN, a DNN layer is represented by 1) a set of global basis vectors that are common for all layers and form a basis system and 2) a set of quantized controlling coefficients specific for each layer that define how the basis vectors are combined to represent the kernels. Such approach requires a novel training method that takes into account the new weight representation. The training approach should conduct the following tasks: 1) how the initial basis vectors are chosen; 2) how to overcome the 
difficulty to optimize the interdependent global basis vectors and the control coefficients together.

As explained in Section~\ref{section:representation}, the set of the global basis vectors to span a vector space should, mathematically, be linearly independent \cite{principles}. Accordingly, the basis vectors in the BasisN framework are initialized to be a set of random orthogonal vectors. The vector space dimension is set based on the size of the RRAM crossbar. For example, if the given RRAM crossbars have the size $256 \times 256$, the vector dimension is set to 256. The control coefficients for each kernel are initialized randomly.

\begin{algorithm}[t]
\caption{BasisN alternating training.}\label{alg:alternating}
\Input{DNN with set of layers $\Gamma$, control coefficients and biases $\Theta= \{C^\gamma,bias^\gamma\}_{\gamma \in \Gamma}$, the set of trainable parameters $\mathcal{T}$, global basis vectors shared between all layers in all crossbars $B=\{b_1, b_2,..., b_d\}$ with d as the size of the crossbars and the dimension of the basis system, loss function $\mathcal{L}$, coefficients learning rate $\eta_{coeffs}$, basis learning rate $\eta_{basis}$, number of training epochs $epochs$, number of coefficients training epochs per alternating cycles $t_{coeffs}$, and number of basis training epochs per alternating cycles $t_{basis}$}

\For {$t_i = 1$ to $epochs$}{
    \uIf{$(t_i \% (t_{coeffs}+ t_{basis}) ==t_{coeffs}+1)$} {
        $B.set\_trainable('True')$\\
        $\eta \gets \eta_{basis}$\\
        \For {$\gamma \in \Gamma$} {
            $C^\gamma.set\_trainable('False')$
        }
        
    }
    \ElseIf {$(t_i \% (t_{coeffs}+ t_{basis}) ==1)$}{
        $B.set\_trainable('False')$\\
        $\eta \gets \eta_{coeffs}$\\
        \For {$\gamma \in \Gamma$} {
            $C^\gamma.set\_trainable('True')$
        }
    }
    Evaluate $\mathcal{L}(\mathcal{T}_{t_i}), \frac{\partial \mathcal{L}}{\partial \mathcal{T}_{t_i}}$\\
    \tcp{$\mathcal{T}_{t_i}$ is the set of trainable parameters at $t_i$} \tcp{either the global basis B} 
    \tcp{or layers' coefficients $C^\gamma,{ \gamma \in \Gamma}$}
    
    $\mathcal{T}_{t_i+1} \gets \mathcal{T}_{t_i} - \eta \frac{\partial \mathcal{L}}{\partial \mathcal{T}_{t_i}}$

}
\end{algorithm}

A difficulty arises when the basis vectors and the control coefficients for kernels are trained together because they are coupled and interdependent \cite{boyd2004convex, nocedal1999numerical}. Therefore, the changes in one variable affect the behavior or performance of the other. Optimizing them simultaneously can lead to conflicts or trade-offs that make it challenging to find an optimum. A method to solve the variables coupling problem is alternating optimization \cite{bezdek2002some, bezdek2003convergence}. The training process in BasisN with this method is shown in Algorithm~\ref{alg:alternating}. 
%
The BasisN training approach alternates between optimizing the control coefficients while keeping the global basis vectors fixed (untrainable) and then switches to fine-tuning the global basis vectors while keeping the control coefficients untrainable. Such cycles of alternating repeat every $t_{coeffs}+ t_{basis}$ epochs, as Algorithm~\ref{alg:alternating} shows. The learning rate and epochs per cycle for the global basis vectors and the control coefficients are set to be $\eta_{basis} << \eta_{coeffs}$ and $t_{basis}<< t_{coeffs}$ 
to avoid severe deviation of the basis from the initial orthogonality condition.


\subsection{Adaptability of BasisN for training from scratch and fine-tuning pre-trained DNNs}

The BasisN training, presented in Section~\ref{sec_training}, is used to train DNN models from scratch without any knowledge from a pre-trained model. However, BasisN training can be adapted to fine-tune a pre-trained DNN  without the need of excessive training epochs. 

The difference between training from scratch as in Section~\ref{sec_training} and fine-tuning is the initialization of the control coefficients of the kernels. Instead of randomly initializing the control coefficients, the control coefficients are initialized to the values that minimize the distance between the original pre-trained kernels and the kernels' representation as combinations of the basis vectors, as described in the following. 



As discussed in Section~\ref{section:representation} and shown in Figure~\ref{fig:method}, a subkernel should be expressed as $\mathbf{k_{i,j}}
=\mathcal{C}_{i,j}\cdot \mathbf{B}$. For a pretrained model, $\mathbf{k_{i,j}}$ is initialized with the kernel values normally trained without considering decomposition.
The basis vectors are still initialized to be orthogonal to each other. The initial coefficients $\mathcal{C}_{i,j}$ for fine-tuning are obtained as $\mathcal{C}_{i,j} =quantize(\mathbf{k_{i,j}} \times \mathbf{B}^{-1}, number\_of\_bits)$, where $\mathbf{B}^{-1}$ is the inverse of the basis matrix and the orthogonality condition of the initialization guarantees the invertibility of $B$, accordingly. $quantize()$ is a quantization function to convert values in $\mathcal{C}_{i,j}$ to the set of allowed values for the coefficients based on the pre-determined number of bits. The DNN model is then fine-tuned using the same alternating training approach as shown in 
Algorithm~\ref{alg:alternating}
to restore the accuracy with fewer epochs than training from scratch.

\subsection{Contest-aware regularization to increase parallelization}
\begin{figure}[!t]
    \centering
    \includegraphics[width=\linewidth]{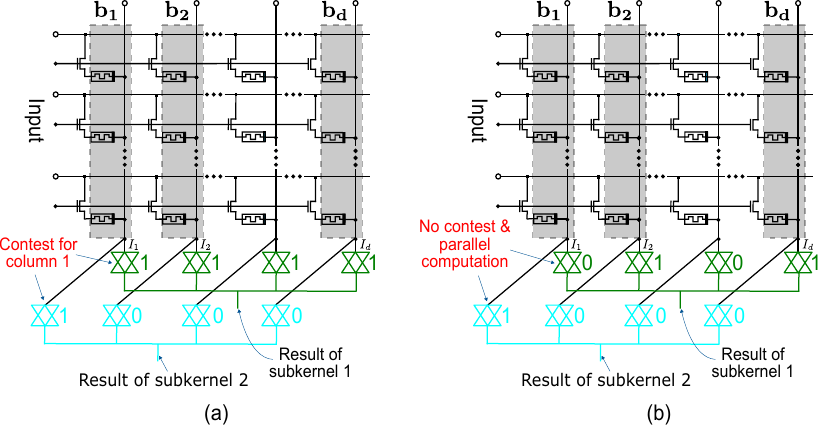}
    \caption{Basis contest between kernels and how it affects parallelization.}
    \label{fig:contest}
\end{figure}

By having only one set of TGs connected to the columns of a crossbar, 
the whole crossbar can only generate the output of one subkernel, because all the columns in the crossbar are combined together with the coefficients. In other words, the whole crossbar is used to implement only one multiplication of a row in the weight matrix with the inputs. Accordingly, the parallelization is degraded. 
To enhance computation efficiency, 
more than one set of TGs are connected in parallel in BasisN as shown in Figure~\ref{fig:contest}. 
However, the number of parallel computations that can be performed at one time step on one crossbar is limited by the contest over some basis vectors.
One basis vector or column in the crossbar must be activated or used by only one TG in all the parallel TGs, i.e., a basis vector should only be activated by one kernel; otherwise, the flowing current representing the output would be mixed and lead to incorrect computation results.

Figure~\ref{fig:contest} illustrates basis contest scenarios. Figure~\ref{fig:contest} (a) shows two sets of TGs connected to the output of the crossbar. 
They implement two kernels with control coefficients $[1,1,1,1], [1,0,0,0]$, respectively.
These two kernels cannot be executed by the crossbar in parallel, because both of them need the basis vector $\mathbf{b_1}$ and thus contest for the corresponding TG. 
  On the contrary, Figure~\ref{fig:contest} (b) shows two kernels
  with control coefficients $[0,1,0,1], [1,0,0,0]$, which can be executed in parallel on the same crossbar without basis contest, because there is no overlap in the TGs.



To allow more kernels to be processed on the same crossbar, 
a regularization term is added to the loss function during training to reduce contest between kernels over basis vectors, i.e., to reduce the number of usages of each basis vector to allow for a higher parallelization. 
The number of `1's in the control coefficients
for each bit significance 
defines how many kernels use the basis vector at that bit significance. 
Hence, the regularization term penalizes the sum of the bits with the value `1' in all coefficients and can be expressed in the loss function as:
$Loss = L_{ce} + \beta * \sum_{\gamma\in \Gamma}
\sum_{i=1}^{\mathcal{K}^\gamma}
\sum_{j=1}^{\mathcal{P}^\gamma_i}
\sum_{l=0}^d 
\sum_{n=0}^N 
(binary(C^\gamma_{i,j,l})\&(2^n))/(2^n)$
, where $L_{ce}$ is the classification crossentropy loss, $\beta$ is a hyperparameter defining the significance assigned to the regularization term, $\Gamma$ is the set of the layers, $\mathcal{K}^\gamma$ is the number of kernels in the $ \gamma$-th layer, 
$\mathcal{P}^\gamma_i$ is the number of partitions of subkernels to fit into $d\times d$ crossbars, 
$N$ is the number of bits in a control coefficient, and
$C^\gamma_{i,j,l}$ is the corresponding coefficient. 
$binary()$ denotes the binary representation of a coefficient as bits. The bit-wise \textit{and} operation (\&) and division with $2^n$ extracts the $n$-th bit of the control coefficient.

\section{Experimental results}\label{sec:results}
\setlength{\tabcolsep}{2.5pt}

\begin{table*}

  \small
    \centering
    \caption{Experimental results of BasisN.}
    \begin{tabular}{cccccccccccc}
        \hline
         \multirow{5}{*}{Network-Dataset} &  \multicolumn{4}{c}{Software Baseline \& Literature} & 
           \multicolumn{7}{c}{BasisN}  \\
         \cline{2-5} \cline{7-12} 
          & \multirow{2}{*}{Software}& \#Weight& \#Available &Ratio of  &&\multirow{3}{*}{Accuracy} & Ratio of& Ratio of&  Ratio of & Ratio of&{Crossbar}\\
          
          &\multirow{2}{*}{accuracy} &fixed&crossbars&\#crossbars &&&inference&inference&energy-delay&energy-delay&area\\


          &&crossbars&on chip \cite{wan2022compute}&in \cite{wan2022compute}&&&cycles to \cite{merced2016repeatable}&cycles to \cite{chen2023novel}&product to \cite{merced2016repeatable}&product to \cite{chen2023novel}&overhead\\
         
         \hline
         \hline
         \multirow{2}{*}{DenseNet-ImageNet}& \multirow{2}{*}{71.5\%}& 
         \multirow{2}{*}{480} & \multirow{2}{*}{48} &\multirow{2}{*}{0.1} &\multirow{2}{*}{} &
         \multirow{2}{*}{72.71\%} &\multirow{2}{*}{0.113\%}&\multirow{2}{*}{0.946\%}& \multirow{2}{*}{\textcolor{black}{0.075\%}}& \multirow{2}{*}{\textcolor{black}{0.063\%}}&\multirow{2}{*}{6.17\%}\\
         
         \multirow{2}{*}{DenseNet-CIFAR100}& \multirow{2}{*}{84.4\%}& 
         \multirow{2}{*}{460} & \multirow{2}{*}{48} &\multirow{2}{*}{0.1403} & \multirow{2}{*}{} &
         \multirow{2}{*}{84.46\%} &\multirow{2}{*}{0.114\%}&\multirow{2}{*}{0.96\%}& \multirow{2}{*}{\textcolor{black}{0.076\%}}& \multirow{2}{*}{\textcolor{black}{0.64\%}}&\multirow{2}{*}{6.17\%}\\

         \multirow{2}{*}{ResNet-CIFAR100}& \multirow{2}{*}{72.34\%}& 
         \multirow{2}{*}{395} &\multirow{2}{*}{48} &\multirow{2}{*}{0.1215} & \multirow{2}{*}{} &
         \multirow{2}{*}{72.22\%} &\multirow{2}{*}{0.26\%}&\multirow{2}{*}{2.2\%}& \multirow{2}{*}{\textcolor{black}{0.434\%}}& \multirow{2}{*}{\textcolor{black}{3.67\%}}&\multirow{2}{*}{6.17\%}\\

         \\
         \hline
    \end{tabular}
    \label{tab:main_table}
\end{table*}

\begin{figure*}[!t]
\begin{minipage}[t]{0.75\linewidth}
    \centering
    \includegraphics[width=\linewidth]{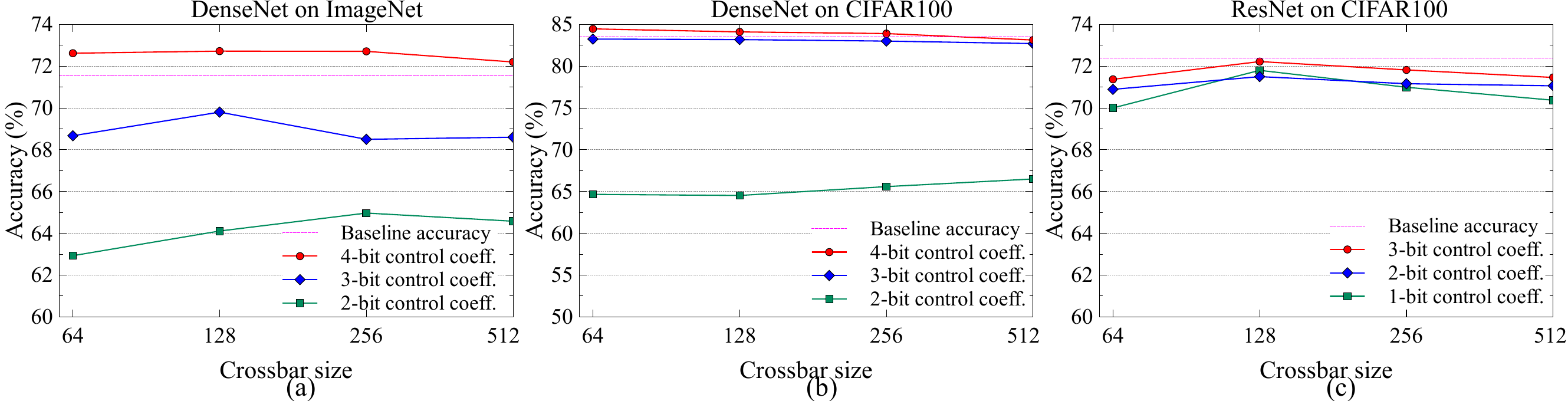}
    \caption{Inference accuracy with respect to the bitwidth of the control coefficient and crossbar size for\\ a) DenseNet-ImageNet b) DenseNet-CIFAR100, and c) ResNet-CIFAR100.}
    \label{fig:Acc_crossDim}
\end{minipage}%
    \hfill%
\begin{minipage}[t]{0.25\linewidth}
    \includegraphics[width=\linewidth]{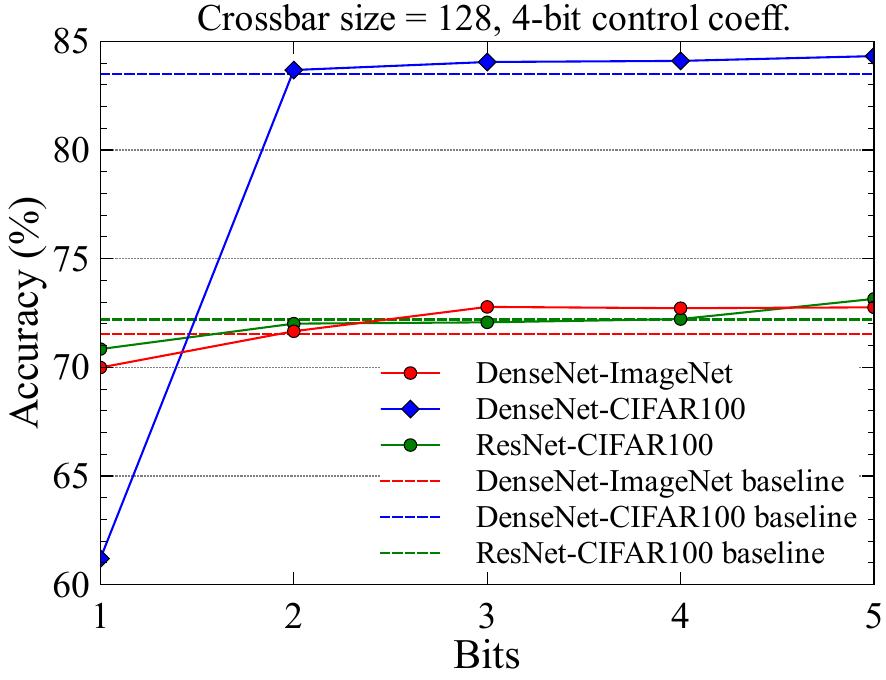}
    \caption{Inference accuracy versus quantization bits per RRAM cell} 
    \label{fig:Acc_RRAMbits}
\end{minipage} 
\end{figure*}

To evaluate the proposed BasisN framework, two DNNs, namely ResNet34 \cite{he2016deep} and DenseNet121 \cite{huang2017densely} were tested on two datasets, namely CIFAR100 \cite{krizhevsky2009learning} and ImageNet \cite{deng2009imagenet}. The filters' coefficients and the basis vectors for ResNet34 were trained from scratch, while the filters' coefficients and the basis vectors in the DenseNet121 were obtained by fine-tuning a pre-trained model for sake of efficiency. 
The DNNs were trained with Nvidia Quadro RTX 6000 GPUs. The area estimation of an RRAM cell and the RRAM crossbars were derived from \cite{yeh2015compact}, \cite{wan2022compute} and used to evaluate the additional overhead incurred by the BasisN framework. The energy estimation in reprogramming an RRAM cell was derived from \cite{zahoor2020resistive} and used to compare the energy consumption with two reprogramming approaches, namely the row-based reprogramming \cite{merced2016repeatable}, and the block-based reprogramming \cite{chen2023novel}. The number of the quantization bits of conductances of RRAM cell in RRAM crossbars representing the quantization of the basis vectors was set to 4.  




Table~\ref{tab:main_table} demonstrates the effectiveness of the BasisN framework in reducing both the number of cycles per inference and the energy consumption per inference for the tested DNNs and datasets. 
The first column shows the tested DNNs and the datasets. The second column shows the baseline software inference accuracy without 
applying the BasisN framework. 
The third column shows the number of crossbars needed to execute each corresponding DNN on 
$256 \times 256$ RRAM crossbars without reprogramming. 
The fourth column shows the 
available number of RRAM crossbars in a manufactured chip \cite{wan2022compute}, which is also considered as our underlying hardware. 
It is clear that 
the number of the crossbars needed is much higher than the available number of RRAM crossbars. 
The fifth column shows the ratio of the available number of crossbars to the number of crossbars needed, which demonstrates the problem of the inability to avoid reprogramming. 

The results of the BasisN framework are shown in the second part of Table~\ref{tab:main_table}. 
The sixth column shows the inference accuracy with the proposed framework, which is similar to the baseline accuracy. 
The seventh and eighth columns show the ratio of the number of cycles per inference needed by BasisN to that required by row-based programming strategy \cite{merced2016repeatable} and block-based programming \cite{chen2023novel}. 
In evaluating these results, the number of available crossbars is 48 and their dimension is $256 \times 256$ as in \cite{wan2022compute}. Accordingly,
reprogramming is inevitable for the techniques from literature. However, BasisN avoids the need to reprogram the crossbars, so that the number of inference cycles was reduced to less than 3\% compared with the two programming strategies, indicating the number of inference cycles can be reduced by more than 97\%. 



The ninth and tenth columns in Table~\ref{tab:main_table} show the ratio of the energy-delay product per inference of BasisN to the energy-delay products in the row-based programming  \cite{merced2016repeatable}, and 
the block-based programming \cite{chen2023novel} of RRAM crossbars. 
BasisN needs no reprogramming of the crossbars, and only the loading of the control coefficients bits causes a small amount of energy dissipation. Accordingly, compared with the previous programming techniques \cite{merced2016repeatable}, \cite{chen2023novel}, the energy-delay product is further reduced to much less than 1\% of the energy-delay product achieved by the previous programming strategies in most of the test cases.
The last column shows the area overhead of BasisN incurred by additional transmission gates. The overhead is evaluated as percentage to the area of the RRAM crossbars. The area overhead is marginal since the RRAM crossbars form a small portion of the total chip area, namely 12\% \cite{wan2022compute}. If the overhead is computed as ratio to the total chip area, it would be less than 1\%. 

\subsection{BasisN inference accuracy with respect to the control coefficients' quantization bits and the crossbar size}
\label{subsection:res_coeff_quantization}

Figure~\ref{fig:Acc_crossDim} demonstrates the inference accuracy of the BasisN framework with respect to two variables for the three benchmarks. The two variables are the control coefficient quantization bits (shown as different curves) 
and the RRAM crossbar size (x-axis). 
As Figure~\ref{fig:Acc_crossDim} shows, the inference accuracy is affected by the number of bits used to represent the control coefficients. 
Besides, the influence of such parameters on the inference accuracy is different for different test cases. 
For DenseNet-CIFAR100 and DenseNet-ImageNet, the highest accuracy is obtained at 4-bit quantization bits matching the software accuracy. 
The control coefficients can be quantized to 3-bits with marginal accuracy loss less than $1\%$ for DenseNet-CIFAR100 and less than $2\%$ for DenseNet-ImageNet. 
For ResNet-CIFAR100, the best performance was obtained at 3-bit quantization. The control coefficients could be quantized to 1-bit with an accuracy loss less than $ 1\%$. 

Figure~\ref{fig:Acc_crossDim} also shows that the inference accuracy is slightly affected by the crossbar size.
For the three benchmarks, a slight accuracy loss, less than $1-2\%$, is noticed at the large crossbar size $512 \times 512$. 
The accuracy loss can be explained with the reduced granularity for very large crossbar sizes. One basis vector is longer and then corresponds to a larger portion of one kernel being controlled by the same coefficients. However, such large crossbars are not practical due to several additional problems such as line resistance and fabrication problems, and are not present in literature. 

\subsection{BasisN inference accuracy with respect to RRAM cells' quantization bits for the basis vectors}
Figure~\ref{fig:Acc_RRAMbits} demonstrates the inference accuracy of the BasisN framework with respect to the number of quantization bits per RRAM cell for the three benchmarks. The crossbar size was  set to be $128 \times 128$ and the control coefficients' quantization was set to 4 bits. 
As Figure~\ref{fig:Acc_RRAMbits} shows, the inference accuracy is robust against the low bit quantization of RRAM cells. For all the three benchmarks, the RRAM cells for the basis vectors could be quantized to as low as 2 bits with no degradation in the inference accuracy compared with the baseline software accuracy represented by the horizontal dashed lines. 

According to \cite{stathopoulos2017multibit,li_analogue_2018}, with complex programming schemes, the max number of bits that can be programmed to an RRAM cell is 6 bits. The robustness of the inference accuracy of BasisN against low-bit width quantization of the RRAM cells relaxes the complexity of the programming approach needed to program the basis values to the RRAM crossbars. The robustness of the inference accuracy against the RRAM cells' qunatization comes from the fact that the information of a DNN layer's kernel is split between the basis vectors stored in the crossbar and the control coefficients.

\subsection{BasisN inference cycles and speedup ablation study}
\label{res:inference}
\begin{figure}
    \centering
    \includegraphics[width=\linewidth]{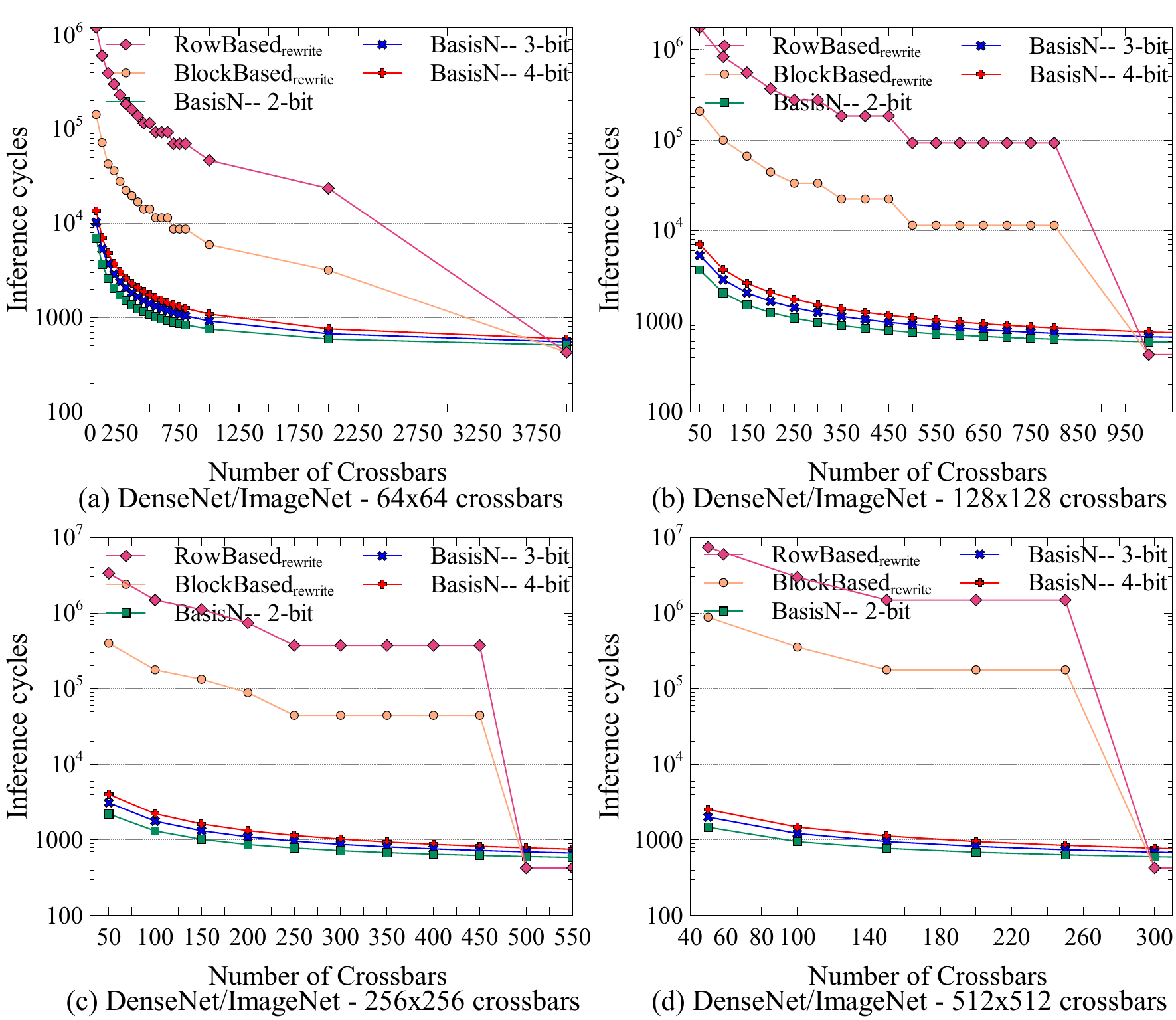}
    \caption{DenseNet-ImageNet number of cycles per inference against the number of on-chip available RRAM crossbars of size a) $64 \times 64$, b) $128 \times 128$, c) $256 \times 256$, and d) $512 \times 512$ for BasisN with different quantization of the control coefficients and comparison with \cite{merced2016repeatable} and \cite{chen2023novel} reprogramming.}
    \label{fig:Cycles_crossbars_densenet_imagenet}
\end{figure}

\begin{figure}
    \centering
    \includegraphics[width=\linewidth]{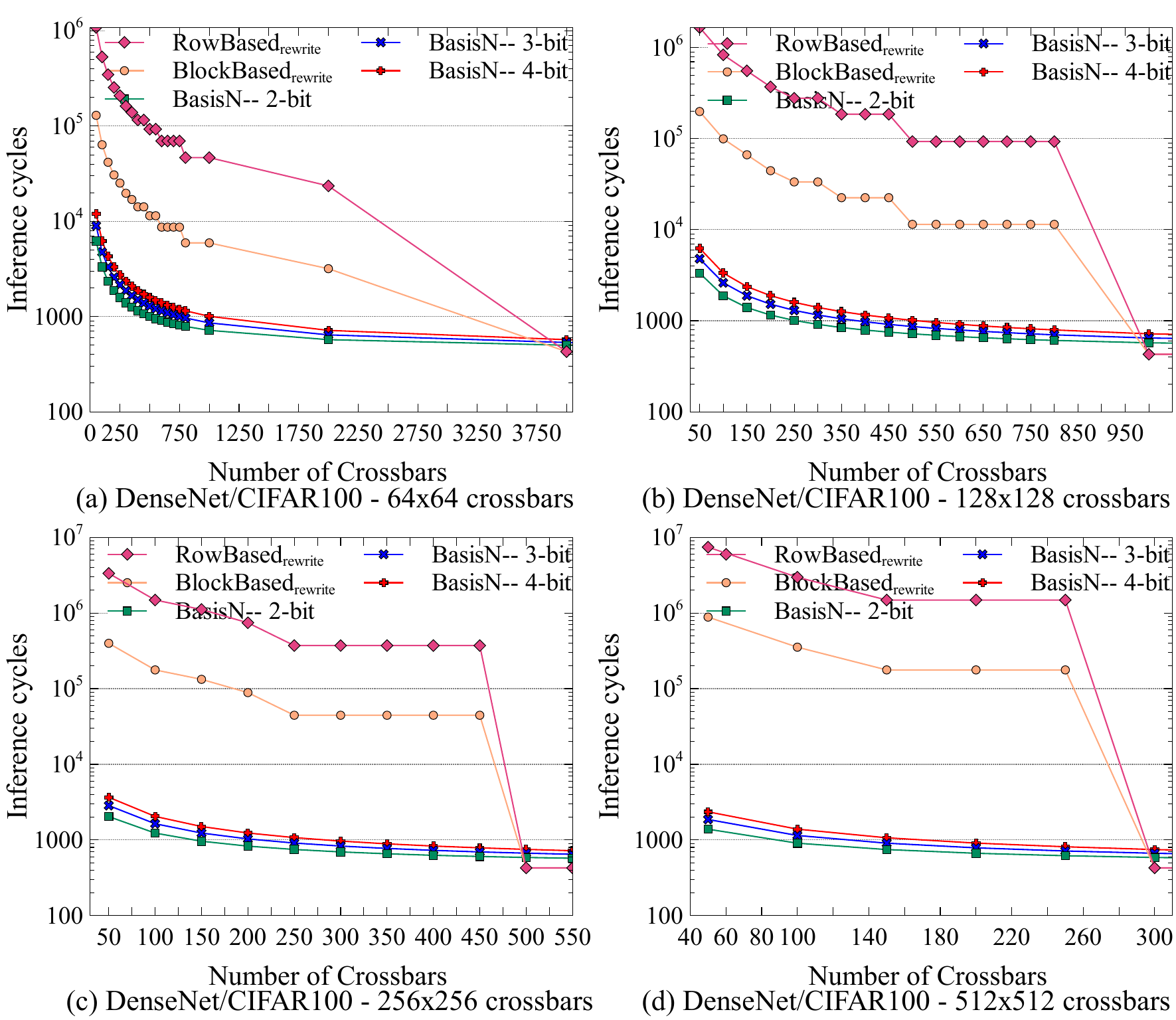}
    \caption{DenseNet-CIFAR100 number of cycles per inference against the number of on-chip available RRAM crossbars of size a) $64 \times 64$, b) $128 \times 128$, c) $256 \times 256$, and d) $512 \times 512$ for BasisN with different quantization of the control coefficients and comparison with \cite{merced2016repeatable} and \cite{chen2023novel} reprogramming.}
    \label{fig:Cycles_crossbars_densenet_cifar100}
\end{figure}

\begin{figure}
    \centering
    \includegraphics[width=\linewidth]{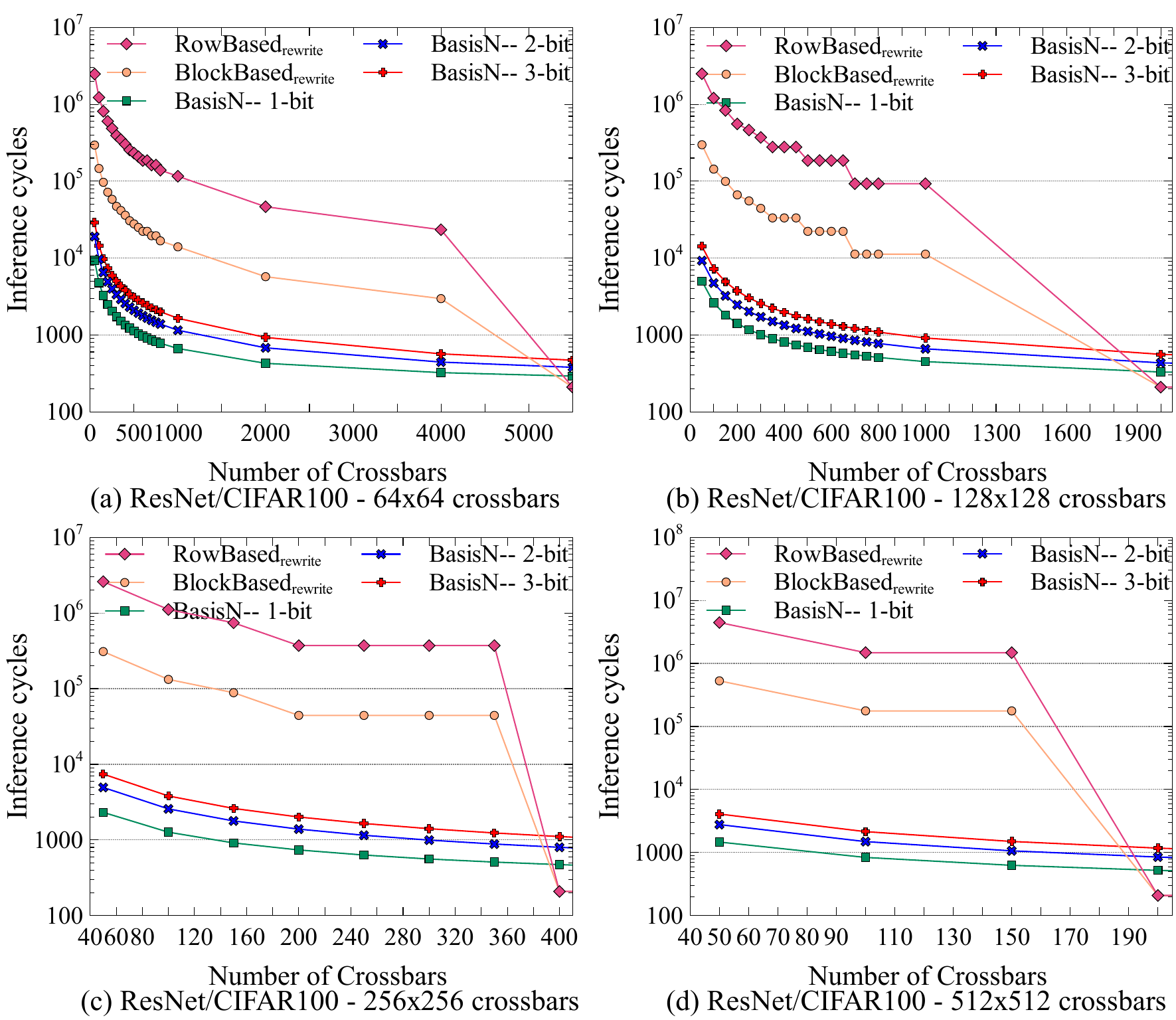}
    \caption{ResNet-CIFAR100 number of cycles per inference against the number of on-chip available RRAM crossbars of size a) $64 \times 64$, b) $128 \times 128$, c) $256 \times 256$, and d) $512 \times 512$ for BasisN with different quantization of the control coefficients and comparison with \cite{merced2016repeatable} and \cite{chen2023novel} reprogramming.}
    \label{fig:Cycles_crossbars_resnet}
\end{figure}

To demonstrate the reduction of inference cycles of the proposed BasisN framework compared with the previous programming strategies, we evaluated the numbers of inference cycles with different number of crossbars and different crossbar sizes. 
Figures~\ref{fig:Cycles_crossbars_densenet_imagenet}, \ref{fig:Cycles_crossbars_densenet_cifar100}, and \ref{fig:Cycles_crossbars_resnet} 
show the comparison results. In such figures, 
 the y-axis represents the number of cycles per inference and the x-axis represents a sweep of the number of available RRAM crossbars on a chip. In subfigures (a), (b), (c), and (d), the corresponding sizes of the RRAM crossbars are $64 \times 64$, $128 \times 128$, $256 \times 256$, and $512 \times 512$, respectively. In each subfigure the number of cycles per inference is plotted for row-based reprogramming \cite{merced2016repeatable}, block-based reprogramming \cite{chen2023novel} and BasisN framework. 
 Besides, 
 different quantization bits for the control coefficients in BasisN were also considered and illustrated. 

Figures \ref{fig:Cycles_crossbars_densenet_imagenet}, \ref{fig:Cycles_crossbars_densenet_cifar100}, and \ref{fig:Cycles_crossbars_resnet} show that, for BasisN, the number of the cycles per inference is dependent on the bit-width of the control coefficients. The higher the bit-width is, the larger the number of the inference cycles becomes. For example in Subfigure~\ref{fig:Cycles_crossbars_densenet_imagenet} (c) the red curve representing inference cycles at a coefficient control quantization of 4 bits is higher than the blue curve with quantization of 3 bits.

According to Figures \ref{fig:Cycles_crossbars_densenet_imagenet}, \ref{fig:Cycles_crossbars_densenet_cifar100}, and \ref{fig:Cycles_crossbars_resnet} , BasisN can reduce the number of inference cycles significantly compared  with row-based \cite{merced2016repeatable}  and block-based \cite{chen2023novel} reprogramming approaches. 
For example, the inference cycles were reduced to less than 10\% of the reprogramming approach \cite{chen2023novel} for all benchmarks under $64 \times 64$ crossbar size and 4-bit control coefficients in subfigures \ref{fig:Cycles_crossbars_densenet_imagenet}(a), \ref{fig:Cycles_crossbars_densenet_cifar100}(a), and \ref{fig:Cycles_crossbars_resnet}(a) . For larger crossbar sizes, BasisN performed even much better and can reduce the inference cycles to $<< 1\%$ of the previous programming technique, which comes from 
the fact that the number of cycles needed to reprogram larger crossbars is larger than for smaller crossbars. 
For example, when the crossbar size is $256 \times 256$ and 48 crossbars were used, 
BasisN reduced the inference cycles to 0.1\% and 0.9\% of that in the reprogramming approaches \cite{merced2016repeatable} and \cite{chen2023novel}, respectively, for the DenseNet-ImageNet and DenseNet-CIFAR100 benchmarks. 

Once the number of the available crossbars becomes large enough to accommodate all the DNN layers' weights without reprogramming, the weight-stationary technique becomes faster than the BasisN framework. For example, the number of needed crossbars to deploy DenseNet-ImageNet benchmark is 480 for a crossbar dimension of $256 \times 256$ without reprogramming.
When the number of crossbars is 500 with the size of $256 \times 256$, as shown 
in subfigure~\ref{fig:Cycles_crossbars_densenet_imagenet} (c), 
the weight stationary approaches have fewer inference cycles since no reprogramming was needed. However, such number of 480 crossbars is unrealistic for RRAM chips. The recent RRAM IMC chips have only about 48 available crossbars on chip \cite{wan2022compute}. 
Similarly, in all other subfigures, weight stationary approaches became faster than BasisN only 
with very large number of crossbars that cannot be fitted on any existing RRAM IMC chips. However, BasisN removed such requirement and could run inferences with any number of available crossbars to achieve fast inference while maintaining the inference accuracy.

\section{Conclusion}
\label{sec:conclusion}
In this paper, we propose the BasisN framework to tackle the problem of the inevitable reprogramming of RRAM crossbars when deploying large DNNs on a limited number of crossbars.  
BasisN introduces a novel representation of the kernels in DNN layers as combinations of global basis vectors shared between all layers with quantized coefficients. These basis vectors are written to crossbars only once and used for the computations of all layers with marginal hardware modification.
A novel training approach was also introduced to train from scratch or fine-tune DNNs that fit BasisN kernel representation. Experimental results demonstrate that cycles per inference and energy-delay product were reduced to below 1\% compared with applying reprogramming on crossbars in processing large-scale DNNs such as DenseNet and ResNet on ImageNet and CIFAR100 datasets, while the training and hardware costs are negligible.



\bibliographystyle{IEEEtran}
\bibliography{IEEEabrv,CONFabrv,bibfile}

\end{document}